\documentclass[twocolumn]{aastex61}

\usepackage{amsmath,graphicx}
 
\received{2017 April 26}
\revised{2017 October 6}
\accepted{2017 October 9}
\submitjournal{ApJ}

\shorttitle{A Misidentified Periodic AGN Behind M31}
\shortauthors{Dorn-Wallenstein, Levesque, \& Ruan}

\begin{document}

\title{A Mote in Andromeda's Disk: \\ a Misidentified Periodic AGN Behind M31}

\correspondingauthor{Trevor Dorn-Wallenstein}
\email{tzdw@uw.edu}

\author[0000-0003-3601-3180]{Trevor Dorn-Wallenstein}
\affiliation{University of Washington Astronomy Department \\
Physics and Astronomy Building, 3910 15th Ave NE  \\
Seattle, WA 98105, USA} 

\author[0000-0003-2184-1581]{Emily M. Levesque}
\affiliation{University of Washington Astronomy Department \\
Physics and Astronomy Building, 3910 15th Ave NE  \\
Seattle, WA 98105, USA}

\author[0000-0001-8665-5523]{John J. Ruan}
\affiliation{University of Washington Astronomy Department \\
Physics and Astronomy Building, 3910 15th Ave NE  \\
Seattle, WA 98105, USA}

\begin{abstract}

We identify an object previously thought to be a star in the disk of M31, J0045+41, as a background $z\approx0.215$ AGN seen through a low-absorption region of M31. We present moderate resolution spectroscopy of J0045+41 obtained using GMOS at Gemini-North. The spectrum contains features attributable to the host galaxy. We model the spectrum to estimate the AGN contribution, from which we estimate the luminosity and virial mass of the central engine. Residuals to our fit reveal a blue-shifted component to the broad H$\alpha$ and H$\beta$ at a relative velocity of $\sim4800$ km s$^{-1}$. We also detect \ion{Na}{1} absorption in the Milky Way restframe. We search for evidence of periodicity using $g$-band photometry from the Palomar Transient Factory and find evidence for multiple periodicities ranging from $\sim80-350$ days. Two of the detected periods are in a 1:4 ratio, which is identical to the predictions of hydrodynamical simulations of binary supermassive black hole systems. If these signals arise due to such a system, J0045+41 is well within the gravitational wave regime. We calculate the time until inspiral due to gravitational radiation, assuming reasonable values of the mass ratio of the two black holes. We discuss the implications of our findings and forthcoming work to identify other such interlopers in the light of upcoming photometric surveys such as the Zwicky Transient Facility (ZTF) or the Large Synoptic Survey Telescope (LSST) projects.

\end{abstract}


\section{Introduction} \label{sec:intro}

Active Galactic Nuclei (AGN) are among the most luminous persistent sources of radiation in the Universe, capable of outshining their host galaxies when in a quasar state. They are hosts to supermassive black holes (SMBHs) and are found throughout the history of the universe from redshift $z\sim7$ onward \citep{mortlock11}. With the advent of surveys like the Sloan Digital Sky Survey (SDSS, \citealt{york00}), the number of cataloged AGN has increased by many orders of magnitude. 

As incredibly powerful sources of ionizing radiation, AGN drive and regulate the evolution of the stars, gas, and dust of their host galaxies. The major merger of two gas-rich galaxies can trigger intense dust production and star formation, while the increased accretion onto the central black hole of one or both galaxies can increase its luminosity, triggering outflows and regulating star formation \citep{sanders88}, leaving behind a massive, gas-poor elliptical remnant. Such mergers appear to be not only frequent, but the primary means by which both SMBHs and galaxies are grown \citep{kauffmann00}. If both galaxies in a merger contain SMBHs, simulations indicate that the black holes themselves can merge over $\sim$Gyr timescales \citep{volonteri03,tremmel17}. At early times, the SMBHs in a merger will appear as dual or offset AGN (depending on the accretion rate of both black holes, \citealt{comerford15}). As their orbits decay, the black holes can form a supermassive black hole binary (SMBHB), which could be observed as an apparently single AGN that displays periodic variability. We present here spectroscopic and time-domain analyses of an AGN behind M31 that has been previously misidentified as a red supergiant, a globular cluster, and an eclipsing binary. We find evidence for the periodic variability of the AGN and discuss the implications of its misidentification in light of forthcoming large photometric surveys.

\subsection{J0045+41}

As a part of a search for red supergiant X-ray binaries --- a still-theoretical class of exotic stellar binary system --- we used the single-epoch photometry of the Local Group Galaxy Survey (LGGS, \citealt{massey03,massey06,massey07}), which covers M31, M33, the Magellanic Clouds and 7 dwarf galaxies in the Local Group, to assemble a statistical sample of Local Group red supergiants (RSGs). We used the method of \citet{massey98} to reduce contamination from the far more prevalent foreground M-dwarfs by taking advantage of the separation of the two populations in $B-V$ vs. $V-R$ space. After creating our sample (and ensuring our results agreed with \citealt{massey09} in M31, where we find 437 candidate RSGs), we searched the {\it Chandra} Source Catalog (CSC, \citealt{evans10}) for X-ray sources within 10'' of the LGGS RSGs. This search yielded one close match.

\begin{figure*}[ht!]
\plotone{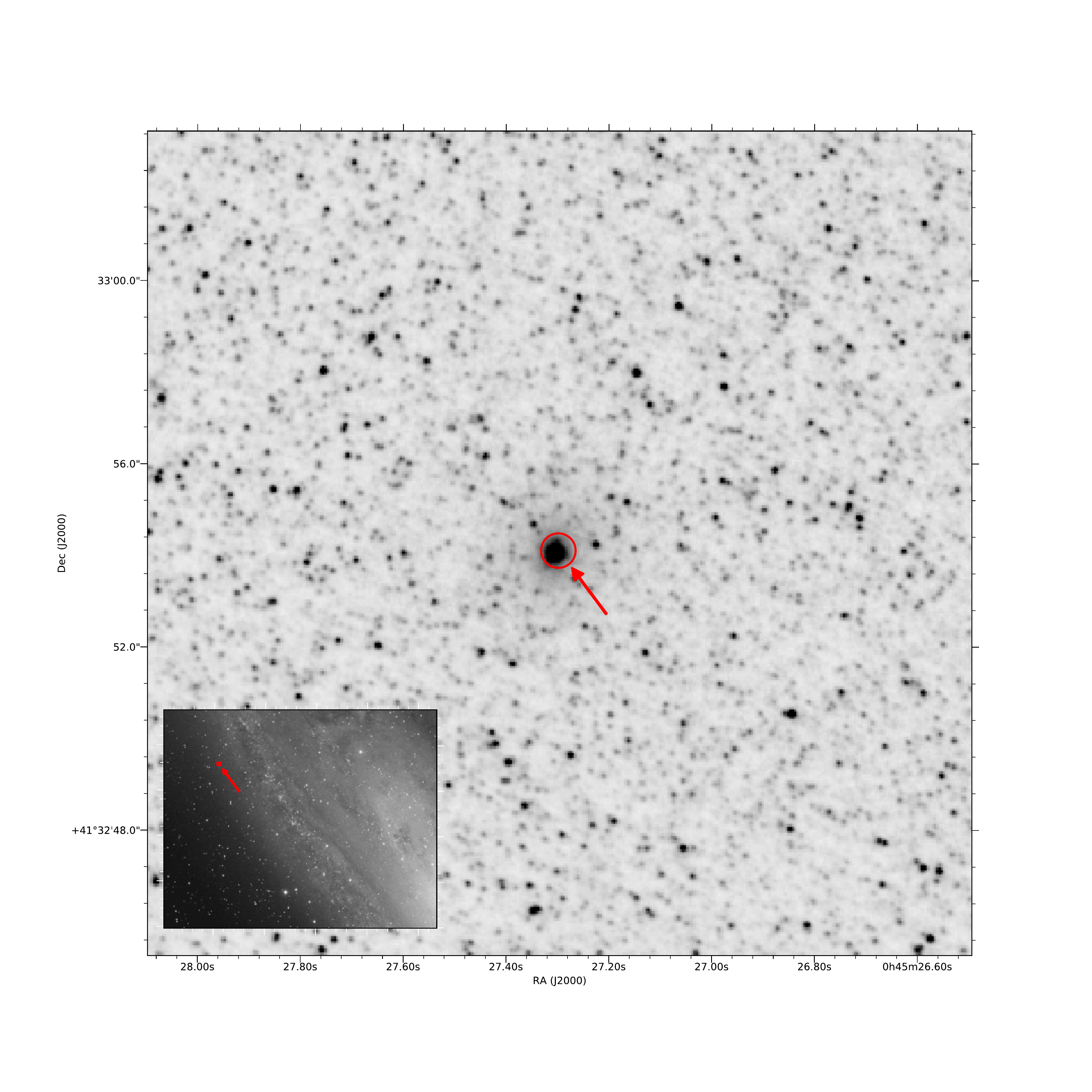}
\caption{{\it F475W} image of J0045+41 from the PHAT survey \citep{dalcanton12}. J0045+41 is the bright object in the center of the image, indicated by the arrow. The location of CXO J004527.3+413255 from the {\it Chandra}PHAT data (Williams et al. {\it in prep}) is indicated by a $0.4''$ positional error circle. Note that the positions of J0045+41 from PHAT and CXO J004527.3+413255 from {\it Chandra}PHAT align even better than the positions from LGGS and CSC. The inset shows the area surrounding J0045+41 on the northeast of M31. The red square indicates the size of the zoomed-in region.\label{fig:PHAT}}
\end{figure*}

LGGS J004527.30+413254.3 ($\alpha = 00^{\rm h}45^{\rm m}27^{\rm s}.30$, $\delta = +41^o32'54''.31$, Figure \ref{fig:PHAT}), which we will refer to as J0045+41 hereafter, is a bright ($V \approx 19.9$) object of previously-unknown nature in the disk of M31. \citet{vilardell06} classified J0045+41 as an eclipsing binary with a period of $\sim76$ days. While the observed variability is of order 1 magnitude in $B$ and $V$, their data are poorly sampled in phase. On the other hand, \citet{kim07} included J0045+41 in a catalog of candidate globular clusters, and it has also been included in catalogs of M31 globular clusters as recently as 2014 \citep{wang14}. The LGGS photometry was consistent with the color and brightness of a typical 12-15 $M_\odot$ RSG in M31, with an inferred effective temperature of $\sim$3500 K and bolometric magnitude of -6.67 (following \citealt{massey09}). However, the best SED fit to photometry from the Panchromatic Hubble Andromeda Treasury (PHAT, \citealt{dalcanton12}) using the Bayesian Extinction And Stellar Tool (BEAST, \citealt{gordon16}) is a 300 M$_{\odot}$, 10$^{\rm 5}$ K ``star'', extincted by $A_V\sim$4 magnitudes, which we exclude as being unphysical. This discrepancy is likely due to the broader wavelength coverage of the PHAT dataset, as well as the fact that the BEAST performs a complete SED fit, whereas our RSG selection criteria are purely based on color and magnitude cuts to select for bright, red objects roughly consistent with the photometric properties of RSGs. Furthermore, the object appears extended in the PHAT images (though its radial profile appears similar to that of other nearby stars; see Figure \ref{fig:PHAT}), implying that J0045+41 may be a background AGN or quasar. Given the angular size of M31 at optical wavelengths ($\sim10$ deg$^2$) and the typical surface density of bright quasars on the sky ($\sim18$ deg$^{-2}$, \citealt{richards02}), we expect $\sim180$ sources in the entirety of M31 to actually be background AGN.

\begin{figure*}[ht!]
\plotone{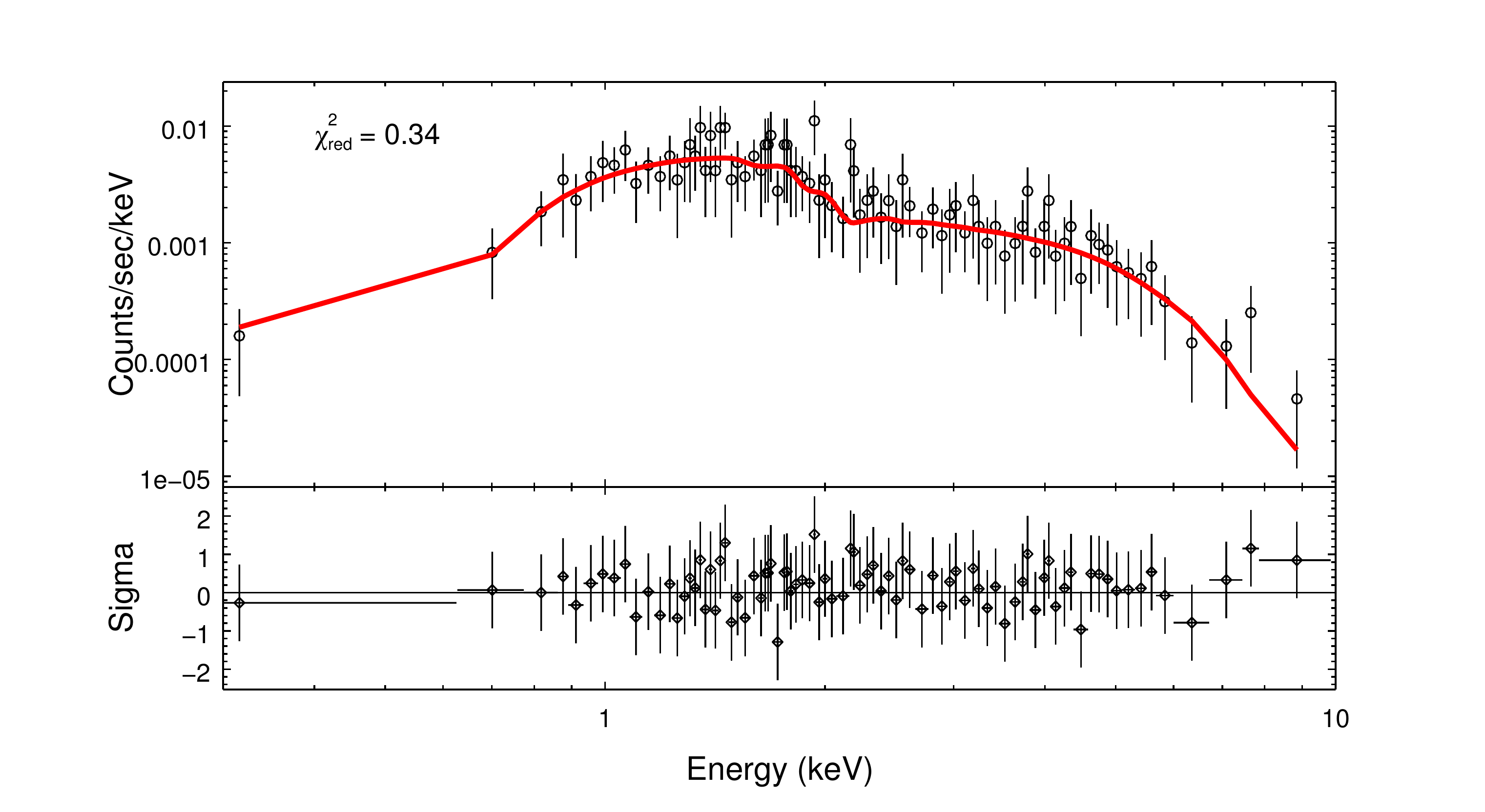}
\caption{{\it Top}: X-ray spectrum of CXO J004527.3+413255, with the best-fit absorbed power law model in red. {\it Bottom}: Fit residuals. \label{fig:xray}}
\end{figure*}

J0045+41 is separated by $\sim1.18^{\prime\prime}$ (4.45 pc at the distance of M31) from an X-ray source in the \citet{evans10} catalog. This source, CXO J004527.3+413255 ($\alpha = 00^{\rm h}45^{\rm m}27^{\rm s}.30$, $\delta = +41^o32'55''.46$), is bright ($F_X = 1.98\times10^{-13}$ erg s$^{-1}$ cm$^{-2}$) and has hardness ratios from \citet{evans10} that are consistent with an unabsorbed X-ray binary or AGN. To confirm this, we fit a spectrum from the publicly available {\it Chandra}PHAT dataset (Obs. ID 17010, \citealt{williams14}) with an absorbed power law model ({\tt xstbabs * powlaw1d}) in {\tt Sherpa} \citep{freeman01}. We use the atomic cross-sections from \citet{verner96}, and abundances from \citet{wilms00}. The spectrum is binned to ensure each bin has a minimum of five counts, and we fit the background-subtracted spectrum from 0.3 to 8 keV. The best-fit ($\chi^2_{red} = 0.34$) model has a neutral H column density $N_H = 1.7\times10^{21}$ cm$^{-2}$ and a power-law slope $\Gamma = 1.5$. The spectrum and fit are shown in Figure \ref{fig:xray}. The value of $N_H$ derived from the fit corresponds to an extinction of $A_V < 1$, which would be surprising if CXO J004527.3+413255 was a background AGN or quasar seen through the disk of M31, as we would expect a significantly higher column density. In addition, using the {\it Chandra}PHAT data, Williams et al. ({\it in prep}) derive improved source locations and positional errors, resulting in a much better alignment between CXO J004527.3+413255 and J0045+41 (see Figure \ref{fig:PHAT})

To conclusively determine the nature of J0045+41, we decided to obtain optical spectrophotometry. We discuss our observations and data reduction in \S\ref{sec:obs}. We present the spectrum, use it to classify J0045+41 as an AGN, identify key features, and analyze it in \S\ref{sec:spec}, and search for evidence of periodicity using archival data in \S\ref{sec:period}. We conclude with a discussion of our results and their implications in \S\ref{sec:discuss}.

\section{Observations and Data Reduction} \label{sec:obs}

We obtained a longslit spectrum of J0045+41 using the Gemini Multi-Object Spectrograph (GMOS) on Gemini-North \citep{hook04}. Four 875 second exposures were taken 2016 July 5 using the \texttt{B600} grating centered on 5000 \AA, and four 600 second exposures were taken 2016 July 9 using the \texttt{R400} grating centered on 7000 \AA, with a blocking filter to remove 2$^{\rm nd}$-order diffraction. Two of each set of exposures were offset by +50 \AA~to fill in the gaps between the three CCDs in GMOS. We followed the standard GMOS-N reduction pipeline using the \texttt{gemini} package in \texttt{IRAF} \citep{gemini16}. Flux calibration was performed using HZ 44 \citep{oke90} as a standard star for both sets of observations. The final reduced spectrum is continuous from $\sim$4000 to $\sim$9100 \AA \ at a resolution of $R \sim 1688\:({\rm blue})/1918\:({\rm red})$. 

\section{Spectrum and Analysis} \label{sec:spec}
\begin{figure*}[ht!]
\plotone{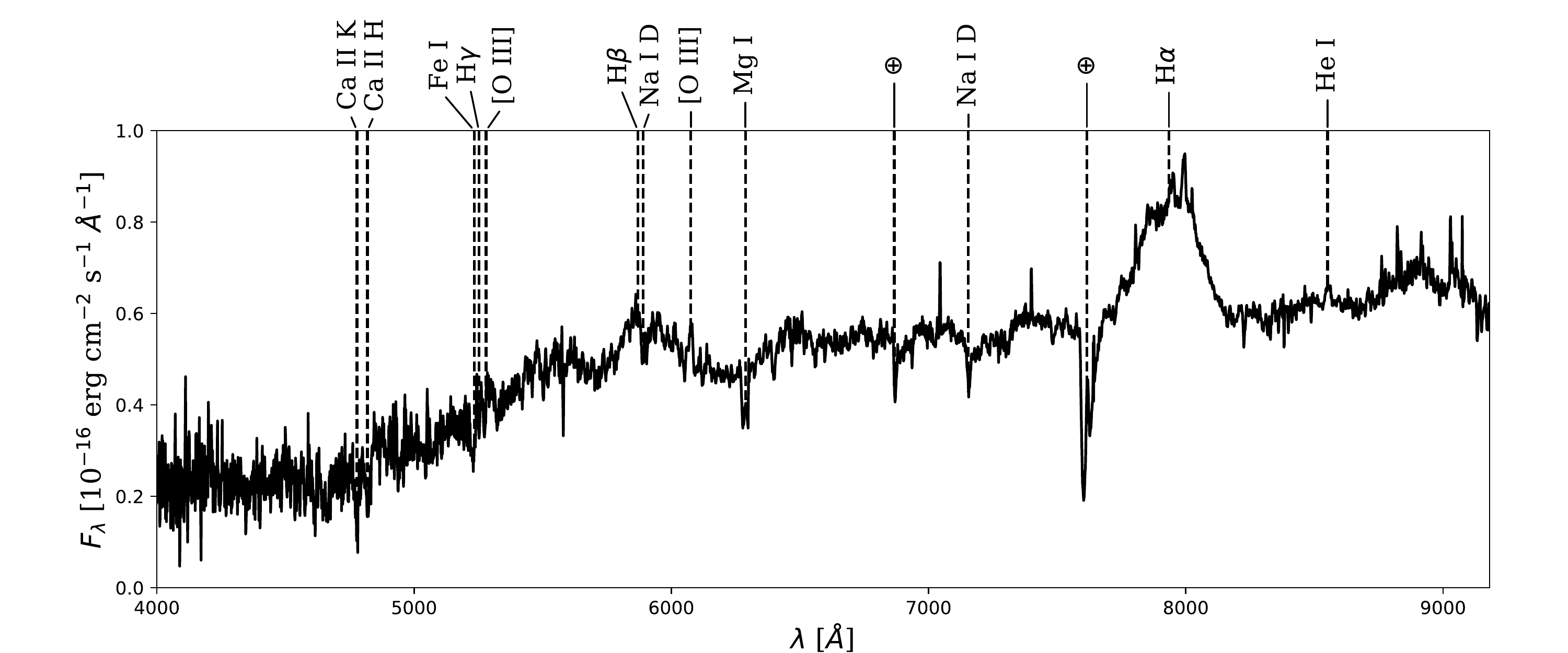}
\caption{GMOS spectrum of J0045+41 with all identified lines labeled. The \ion{Na}{1} feature coincident with H$\beta$ is intrinsic to the Milky Way/M31. The atmospheric O$_2$ A and B bands are marked with $\oplus$.\label{fig:spectrum}}
\end{figure*}

The optical spectrum is shown in Figure \ref{fig:spectrum}. It shows the broad emission lines characteristic of an AGN. We use \ion{Ca}{2} H \& K, the \ion{Fe}{1}/H$\gamma$/[\ion{O}{3}] G band, [\ion{O}{3}] $\lambda5007$, \ion{Mg}{1} $\lambda\lambda$5192,5197, \ion{Na}{1} D, and \ion{He}{1} $\lambda7067$ to determine that J0045+41 is at $z \approx 0.215$. We also detect \ion{Na}{1} D doublet absorption in the rest frame of the Local Group; however our data are not of sufficient resolution to distinguish Milky Way from M31 absorption. Both H$\alpha$ and H$\beta$ are broad, with full widths at half maximum of $\sim10^4$ km s$^{-1}$. The centers of broad H$\alpha$ and H$\beta$ are slightly blueshifted ($z \approx 0.21$) relative to the rest of the spectrum, which may be indicative of an outflow or motion of the central engine relative to the host galaxy.

\begin{figure*}[ht!]
\plotone{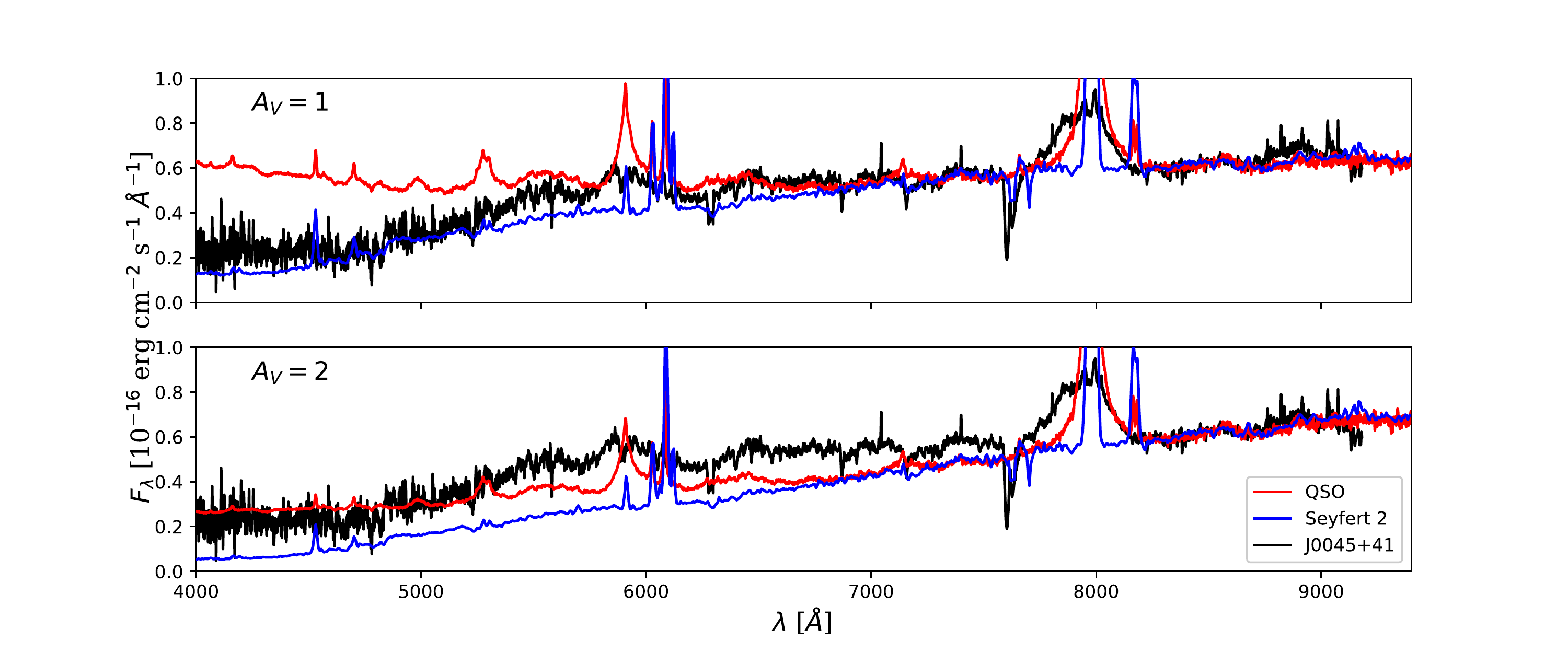}
\caption{{\it Top}: J0045+41 compared with the \citet{vandenberk01} composite quasar spectrum and PySynphot Seyfert 2 spectrum, redshifted to $z = 0.215$, and reddened with $A_V = 1$, the approximate extinction implied by the X-ray spectrum. {\it Bottom}: Same as above, but with $A_V = 2$.\label{fig:compare}}
\end{figure*}

Mistaking a blue AGN for a red star might seem unsurprising given that it is seen through the disk of M31. However, the low amount of extinction implied from the fit to the X-ray spectrum seems inconsistent with an object seen through an entire galactic disk. In Figure \ref{fig:compare}, we show our spectrum of J0045+41 compared with the composite Sloan Digital Sky Survey (SDSS, \citealt{york00}) quasar template spectrum from \citet{vandenberk01} as well as a template Seyfert 2 spectrum from PySynphot (a Python implementation of Synphot destributed by Space Telescope Science Institute, \citealt{lim15}), both redshifted to $z=0.215$ and reddened by 1 (top) and 2 (bottom) magnitudes of extinction in $V$ using a standard \citet{cardelli89} $R_V = 3.1$ extinction law. While hardly a robust fit, this comparison serves to illustrate that either a larger value of extinction is required to reproduce the overall spectral shape of J0045+41 with a pure QSO template or that many of the spectral features --- e.g., the apparent break in spectral slope at $\sim5500$ \AA\ and the presence of strong absorption lines in the spectrum -- are intrinsic to the host galaxy. 

\begin{figure*}[ht!]
\plotone{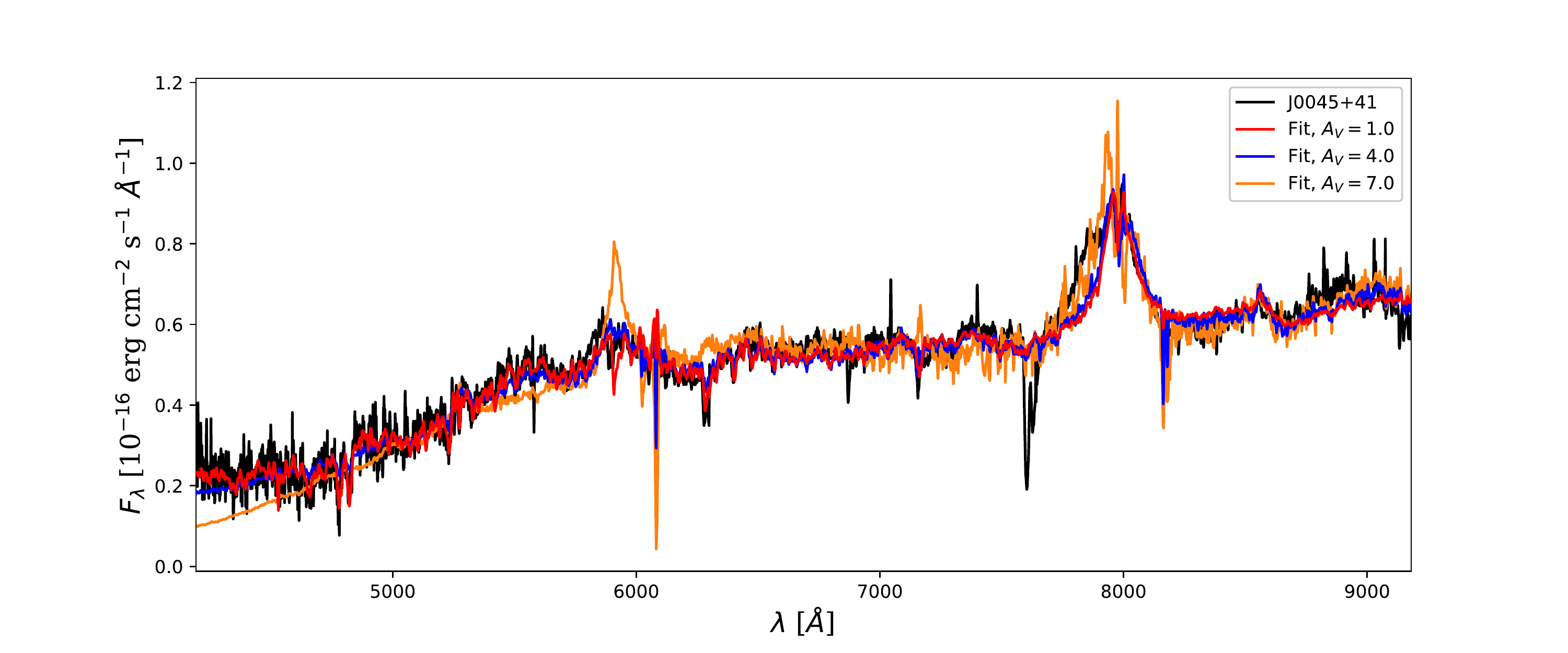}
\caption{Comparison of our fit to the (black) spectrum with the \citet{yip04,yip04b} eigenspectra, after redshifting the basis set to $z = 0.215$, and reddening with a \citet{cardelli89} extinction law for various $A_V$. At larger values of $A_V$, a good fit is impossible, confirming that the extinction through M31 is low at this location. In all cases where a good fit is found, there is good agreement between the model and fit for the locations of most absorption lines, but H$\alpha$ and H$\beta$ both have an excess in the blue. \label{fig:PCA}}
\end{figure*}

To decompose the spectrum into host and AGN spectra, we follow \citet{vandenberk06}. We use the first five galaxy eigenspectra and the first ten QSO eigenspectra derived from a Principal Component Analyses (PCA) of SDSS galaxy and quasar samples \citep{yip04,yip04b} as a set of basis spectra, which we redden using the \citet{cardelli89} extinction law, redshift to $z=0.215$, and fit to our spectrum of J0045+41 as follows. If the measured fluxes are represented by a column vector, $f$, then the residuals between the data and the basis spectra fit is simply
\begin{equation}
E = f - G\cdot c
\end{equation}
where $G$ is a matrix whose columns are the redshifted and reddened basis spectra interpolated to the values of the observed wavelengths in our spectrum and $c$ is a column vector containing the coefficients for each basis spectrum. Taking the errors on each point into account, the scaled residual at each point can be represented by the scalar
\begin{equation}
R = E^T\Sigma^{-1}E
\end{equation}
where $\Sigma$ is the covariance matrix and $E^T$ denotes the matrix transpose. It can be shown that the coefficients that minimize $R$ are given by
\begin{equation}
c = (G^T\Sigma^{-1}G)^{-1}(G^T\Sigma^{-1})\cdot f
\end{equation}
In order to estimate a suitable value of $A_V$ to use when reddening the basis spectra, we redden the spectra with integer values of $0\leq A_V\leq 10$ mag. Some of these fits are shown in Figure \ref{fig:PCA}. While the basis spectra sufficiently fit the spectrum for $0 \leq A_V \leq 2$ mag, at higher values, the basis spectra are unable to reproduce the spectral shape, especially in the blue. Going forward, we adopt $A_V = 1$ mag. \citet{dalcanton15} mapped the dust extinction in M31 at a resolution of 25 pc using the PHAT dataset. They model the probability distribution of $A_V$ in each pixel with a log-normal distribution, parametrized by the median extinction, $\tilde{A}_V$ and the dimensionless width, $\sigma$, such that the mean extinction $\langle A_V\rangle$ is 
\begin{equation}
\langle A_V\rangle = \tilde{A}_Ve^{\sigma^2/2}
\end{equation} 
and the variance in the extinction $\sigma_A^2$ is
\begin{equation}
\sigma_A^2 = \tilde{A}_V^2e^{\sigma^2}(e^{\sigma^2} - 1)
\end{equation}
\citet{dalcanton15} also include the fraction of stars in each pixel that are reddened, $f_{red}$. In the pixel containing J0045+41, $f_{red} = 0.206$, $\tilde{A}_V = 0.72$, and $\sigma = 0.28$. The latter two values correspond to $\langle A_V\rangle = 0.75$, $\sigma_A = 0.21$, consistent with our estimate of $A_V$. Spectral modeling at higher resolution would further constrain the extinction along the particular line of sight towards J0045+41.

\begin{figure*}[ht!]
\plotone{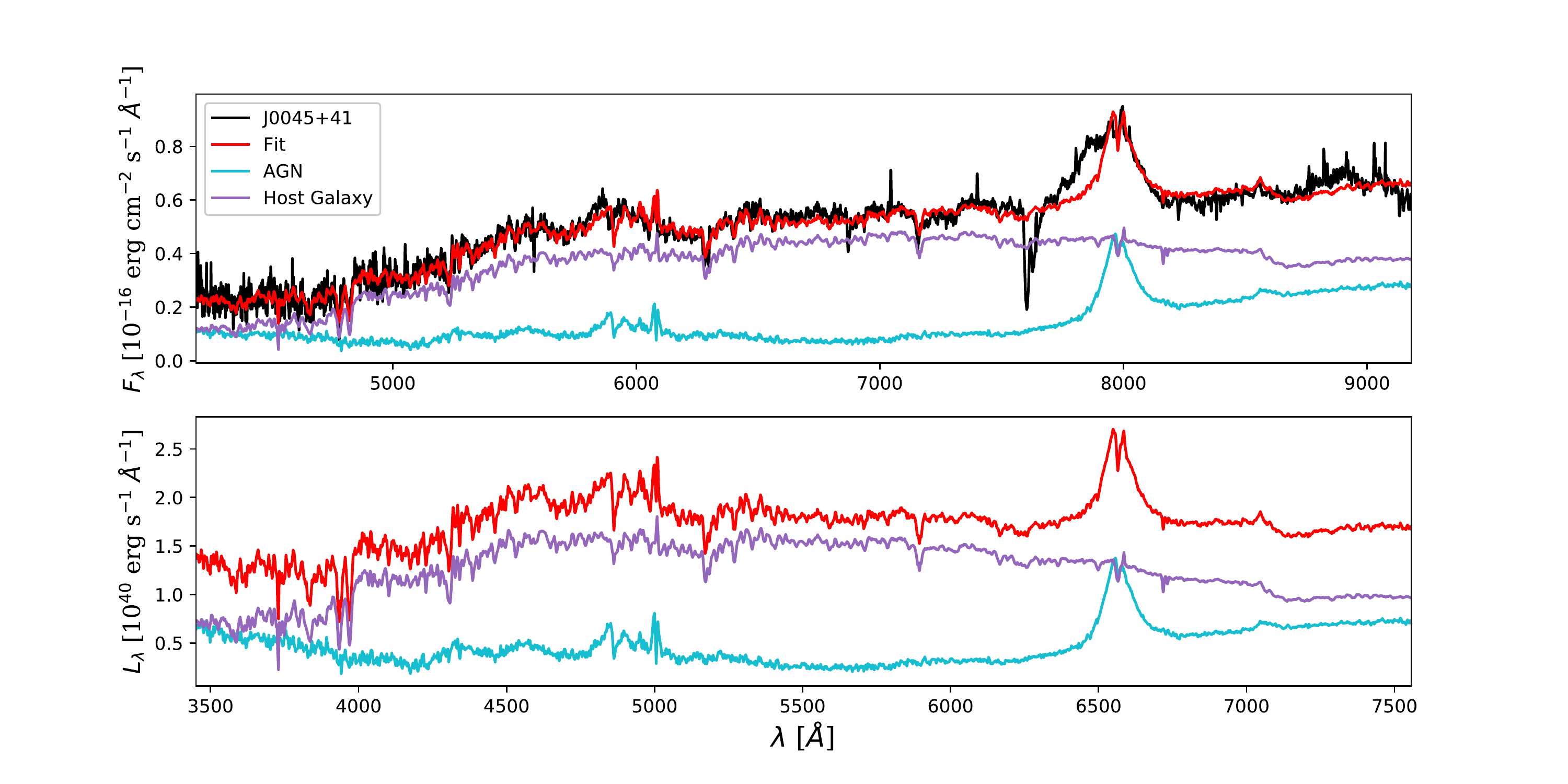}
\caption{{\it Top}: The true spectrum (black) with the PCA fit (red), AGN (cyan) and galaxy (purple) components overlaid. {\it Bottom}: Dereddened rest-frame luminosity spectra of the fit, quasar and galaxy components. \label{fig:fit}}
\end{figure*}

The galaxy and AGN components of this fit are shown in the top panel of Figure \ref{fig:fit}. The bottom panel shows the dereddened rest-frame luminosity spectrum of each component. The luminosity of the underlying AGN component is $L_\lambda = 3.46\times 10^{39}$ erg s${-1}$ \AA$^{-1}$ at 5100 \AA. The derived host galaxy spectrum appears similar to an early type galaxy. This is unsurprising as the hosts of low-luminosity AGN (like J0045+41) tend to be early type \citep{kauffmann03}. If the periodicity (discussed in Section \ref{sec:period}) arises from a SMBH binary formed through the major merger of two late type AGN hosts, it would also be unsurprising that the resulting host is an early type galaxy.

With the underlying contribution to the spectrum from the central engine now known, it is possible to estimate the mass of the SMBH \citep{shen08}. We use the full width at half maximum of H$\beta$ ($1.11\times 10^4$ km s$^{-1}$), the continuum rest frame luminosity from the quasar at 5100 \AA, and the H$\beta$ virial mass estimator coefficients from \citet{mclure04} to calculate $\log(M/M_{\odot}) = 8.30$. We use the bolometric correction from \citet{runnoe12} to calculate the bolometric luminosity, from which we determine the Eddington ratio $\Gamma \equiv L_{bol} / L_{Edd} = 0.007$. This small value for $\Gamma$ may indicate that the accretion flow is radiatively inefficient \citep{casse04}. 

\section{Potential Periodicity} \label{sec:period}

\begin{figure*}[ht!]
\plotone{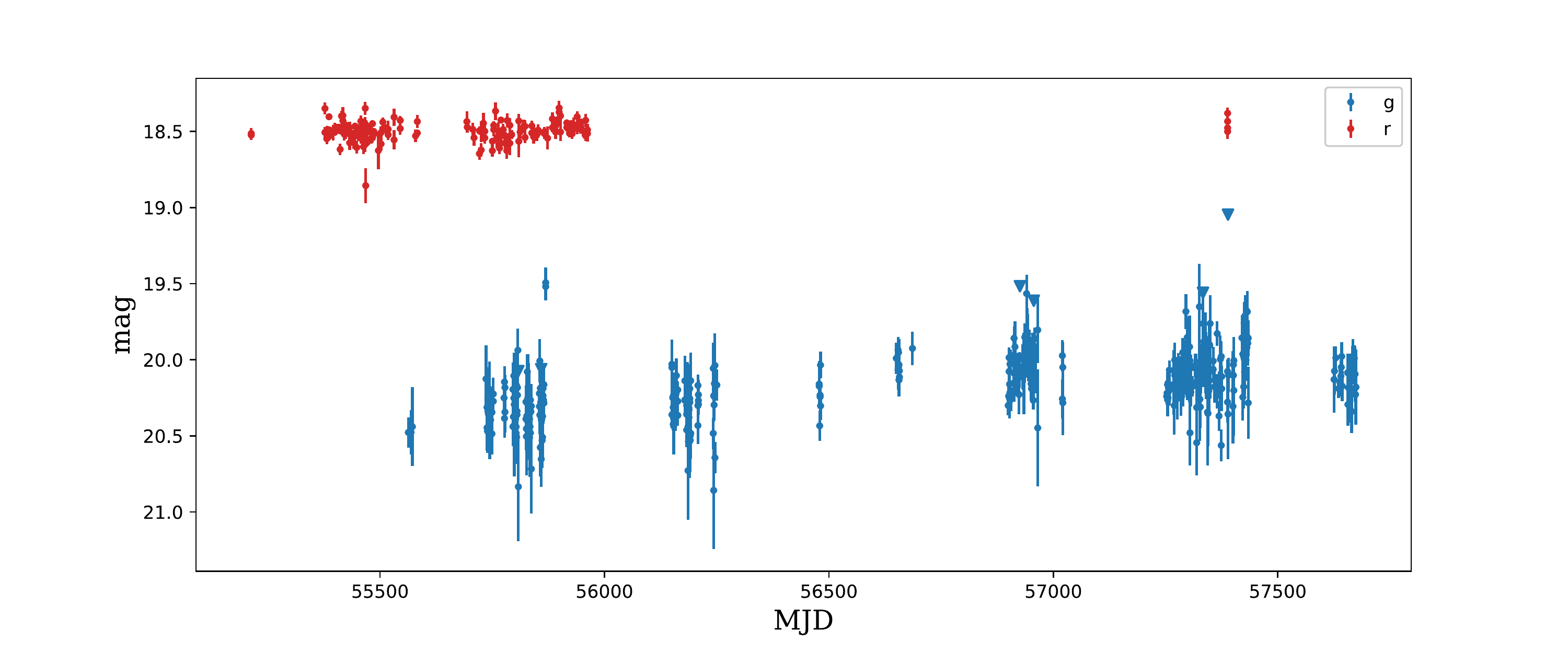}
\caption{Observed PTF lightcurve in $g$ and $r$. Upper limits for any non-detections are shown as downward pointing triangles\label{fig:PTF_lc}}
\end{figure*}

\subsection{Searching for Periodicity Using the {\normalfont Supersmoother} Algorithm}

Though the light curve in \citet{vilardell06} is sparsely sampled, the suggestion of a $\sim$76 day period in J0045+41 prompted further investigation. While continuum emission from AGN is well-known to be stochastically variable due to a variety of phenomena associated with the central engine and surrounding environment, periodicities in the variability have long been predicted as a signature of SMBHBs (e.g., \citealt{bogdanovic08}). A short-period SMBH system would be well within the gravitational wave regime. We investigated the reported periodicity using data from the Palomar Transient Factory (PTF, \citealt{law09}). PTF observed J0045+41 in both $g$ and $r$, though the $g$-band data cover a broader range in time, and thus we focus our analyses solely on those data. These data are shown in Figure \ref{fig:PTF_lc}.

AGN continuum variability is well fit by a damped random walk (DRW) process \citep{kelly09}, described by a characteristic timescale ($\tau$) and long-term rms variability ($\sigma$ or $SF_\infty = \sqrt{2}\sigma$). The power spectral distribution (PSD) of a DRW process \citep{charisi16} is
\begin{equation}
PSD(T) = \frac{4\sigma^2\tau}{1 + 4\pi(\tau/T)^2}
\end{equation}
and the covariance function is
\begin{equation}\label{eq:DRW_cov}
S(\Delta t) = \sigma^2e^{-|\Delta t|/\tau}
\end{equation}
where $\Delta t$ is the time between two observations.

Previous searches for periodicities in AGN lightcurves commonly use Lomb-Scargle periodograms (\citealt{liu16,charisi16,zheng16}). Lomb-Scargle periodograms detect periodicities in irregularly-sampled lightcurves by fitting sinusoids to the data \citep{lomb76,scargle82}. It is important to note that sinusoidal variability is expected if the periodicity arises due to the relativistic Doppler boost of the emission of the secondary component of a steadily-accreting binary (see \citealt{dorazio15}). However, the predicted periodicity from SMBHBs is not necessarily sinusoidal if caused by periodic episodes of accretion (e.g., \citealt{farris15}). Furthermore, \citet{vaughan16} show that the behavior generated by red noise processes can be well fit by a sinusoid over a few `cycles'. Therefore the statistical significance of previously-reported detections using Lomb-Scargle periodogram analysis may be overestimated. 

\begin{figure}[ht!]
\plotone{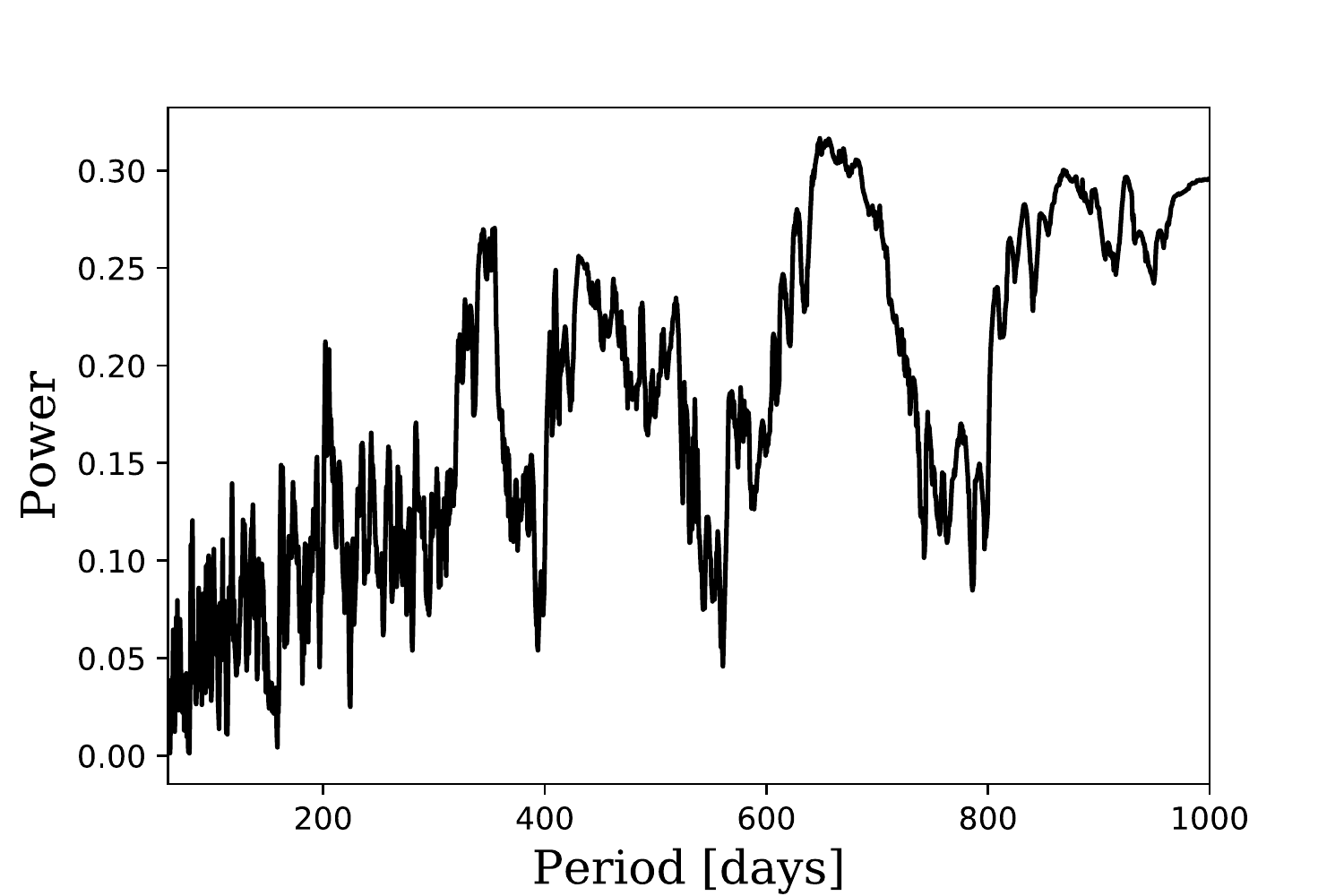}
\caption{{\it Supersmoother} periodogram calculated from the PTF data.\label{fig:super_gram}}
\end{figure}

To provide a robust assessment of periodicities in the lightcurve of J0045+41, we utilize the {\it Supersmoother} algorithm \citep{reimann94}, which uses a non-parametric periodic model to test the strength of signals at various periods. Using the implementation in the {\tt gatspy} Python package \citep{vanderplas15}, we calculate the periodogram of the $g$-band data on a linearly spaced grid of 2000 periods between 60 and 1000 days --- we are unlikely to see periods shorter than 60 days (see \citealt{charisi16}), and our data do not cover more than two cycles of a signal with more than a 1000 day period. The periodogram is shown in Figure \ref{fig:super_gram}. As expected by a DRW signal, the power appears to rise to a constant level at long periods. However, there do appear to be real peaks superimposed onto the expected DRW behavior. 

\subsection{Estimating the Significance of Measured Peaks}

\begin{figure*}[ht!]
\plottwo{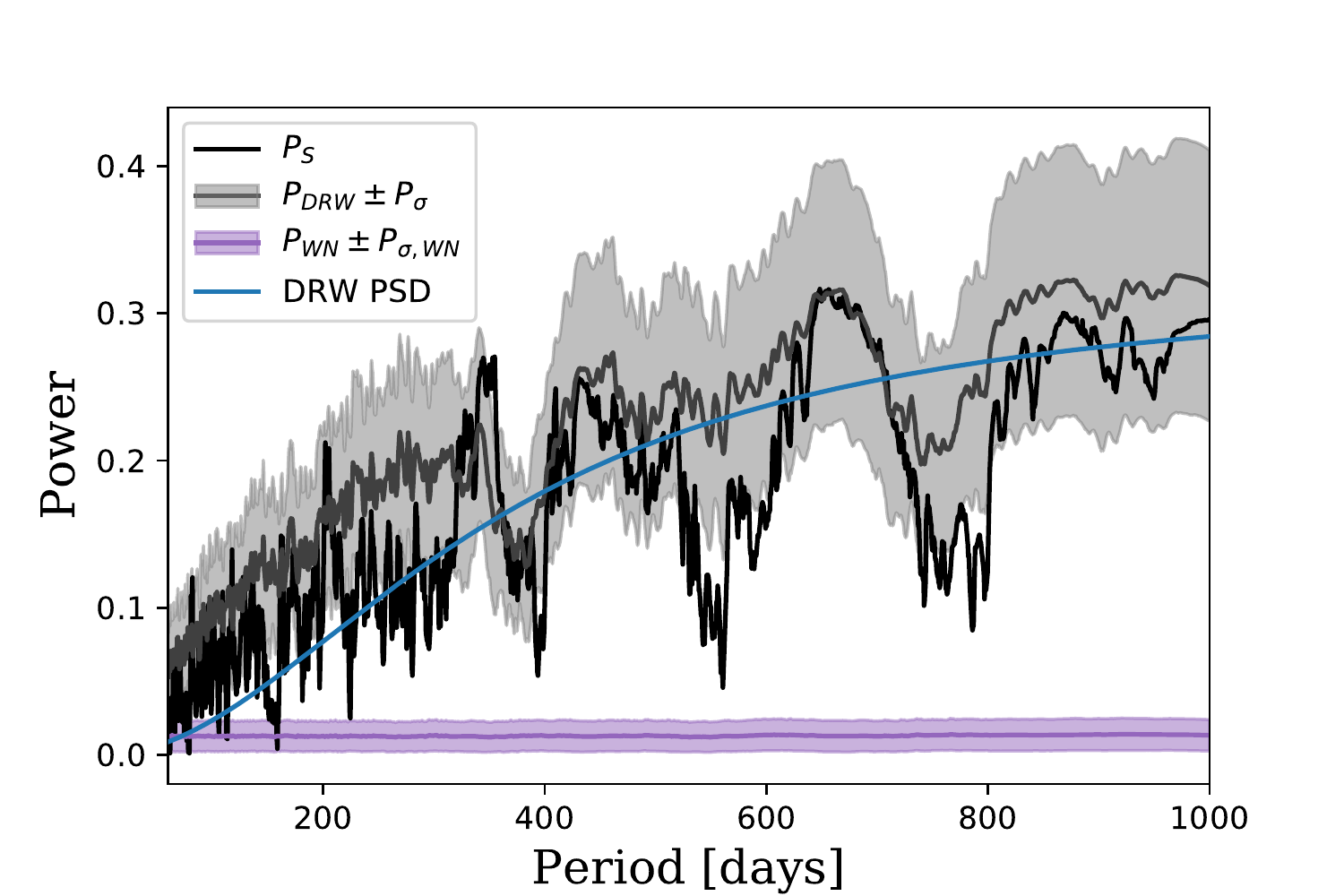}{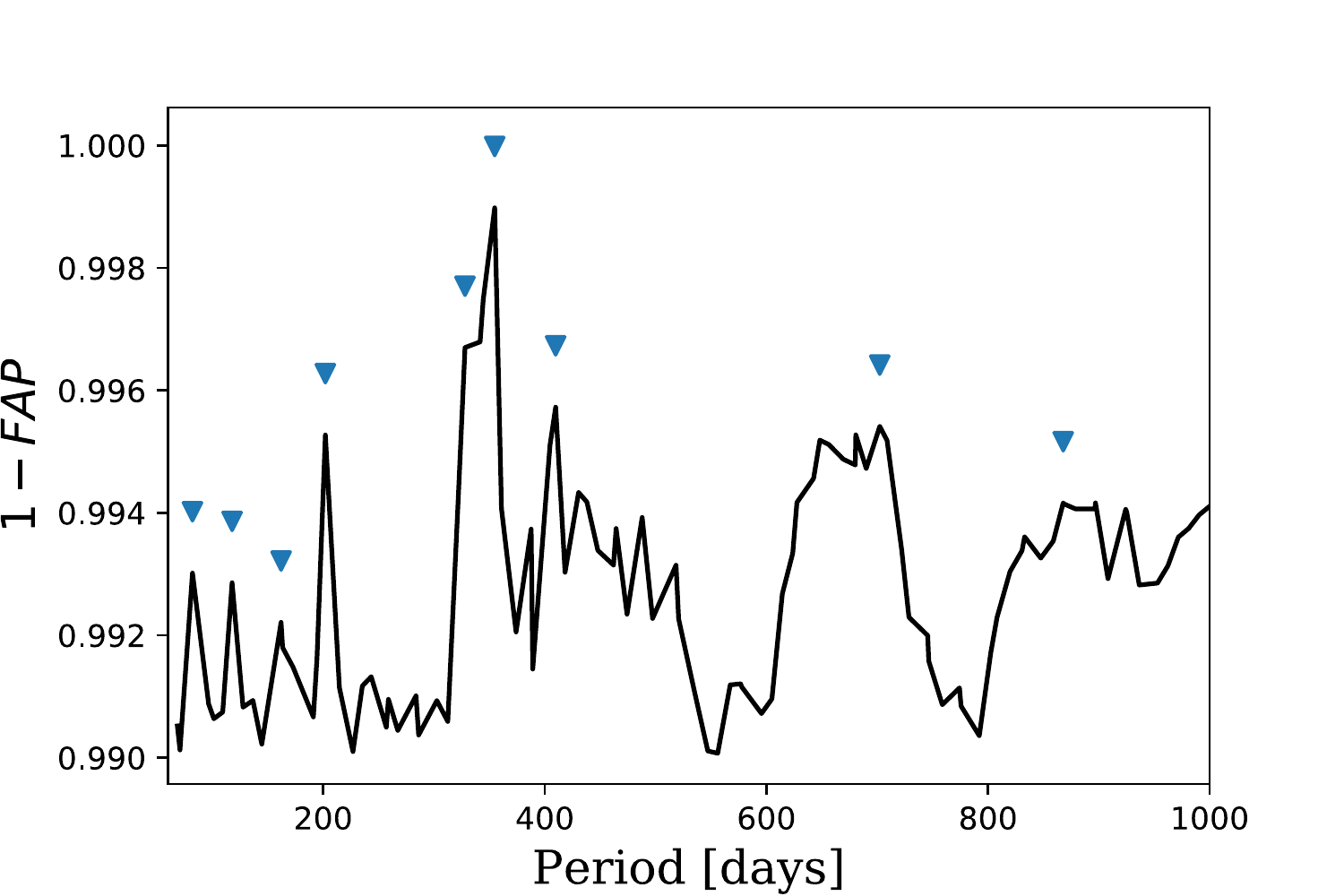}
\caption{{\it Left}: The true periodogram (black) compared to $P_{DRW} \pm P_\sigma$ (gray), $P_{WN} \pm P_{\sigma,WN}$ (purple, described below), and the PSD for a perfectly sampled DRW lightcurve (blue, scaled to roughly match the normalization of the {\it Supersmoother} periodogram). {\it Right}: $1-$FAP vs. $T$, with minima identified with blue triangles. The period with the lowest FAP is at $T\sim$355 days.\label{fig:periodogram}}
\end{figure*}

To check that the measured power of the true signal ($P_S(T)$) is not attributable to a DRW process, we generate simulated DRW lightcurves, following the prescription of \citet{macleod10}, {and compare the distribution of the periodograms of the simulated lightcurves to $P_S(T)$. While it is possible to calculate the DRW parameters, $\sigma$ and $\tau$, from the estimated mass of J0045+41, we choose to instead estimate those parameters by fitting the lightcurve directly, thus incorporating the distribution of possible values. We implement \eqref{eq:DRW_cov} as a kernel function in {\tt celerite} \citep{foremanmackey17}, a Python package for Gaussian process computations, which calculates the likelihood, $\mathcal{L}$, of a DRW with given $\sigma$ and $\tau$:
\begin{equation}
\ln\mathcal{L} = -\frac{1}{2}r^TK^{-1}r - \frac{1}{2}\ln |K| - C
\end{equation}
where $r$ is a vector of the observed data minus the mean, $K$ is the covariance matrix incorporating the photometric errors and the DRW covariance function, and $C$ is a constant proportional to the number of measurements (for a discussion of Gaussian processes and the derivation of this likelihood function, see \citealt{rasmussen06}). We then use {\tt emcee} \citep{foremanmackey13}, an affine-invariant MCMC Python package, to fit for $\sigma$, $\tau$, and the mean magnitude $\langle g\rangle$ by sampling the posterior distribution. We use 32 walkers, and, after discarding 500 burn-in steps, record 3000 samples per walker for a total of 96,000 samples. A corner plot of these samples is shown in Figure \ref{fig:drw_samples}.

\begin{figure}[ht!]
\centering 
\includegraphics[width=0.4\textwidth,bb= 20 20 540 540]{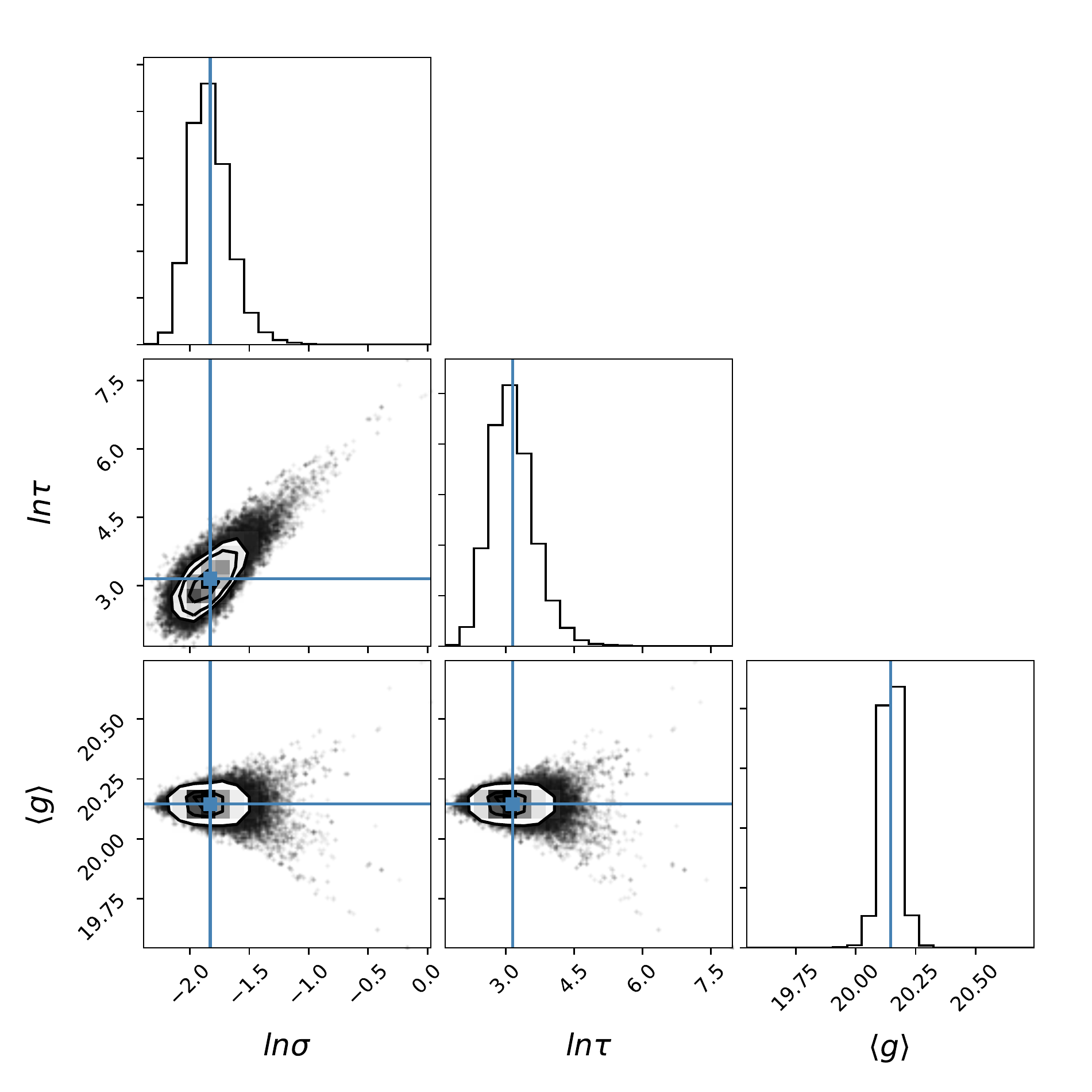}
\caption{Posterior distribution of $\ln\sigma$, $\ln\tau$ and $\langle g\rangle$, sampled by {\tt emcee}\label{fig:drw_samples}}
\end{figure}

Drawing the value of $\sigma$, $\tau$ and $\langle g\rangle$ from the posterior distribution of samples, we generate 96,000 DRW lightcurves.} The lightcurves are sampled at the same times as the PTF observations and have identical photometric errors. The final points in the simulated lightcurve are then drawn from a Gaussian distribution with the magnitude of the raw point as the mean, and standard deviation equal to the photometric error. We then calculate periodograms for each simulated DRW lightcurve on the same grid of periods as $P_S$. The mean ($P_{DRW}$) and standard deviation ($P_{\sigma}$) of the simulated periodograms are plotted as $P_{DRW}\pm P_{\sigma}$ along with $P_S$ and the theoretical DRW PSD with $\sigma = 0.2$, $\tau = 200$ days (scaled to match the values returned by {\it Supersmoother}) for comparison in the left panel of Figure \ref{fig:periodogram}. Much of the structure in the true periodogram is matched by the simulated periodograms, but not in the theoretical PSD. This is likely due to the irregular sampling of the PTF lightcurve, which is reflected in the simulated lightcurves. However, some of the peaks in the true periodogram do not appear in the DRW noise.

To identify periods with power in excess of the DRW noise, we search for peaks in $\sigma = (P_S - P_{DRW})/P_{\sigma}$. $\sigma(T)$ is plotted in the right panel of Figure \ref{fig:periodogram}, with the ten peaks with largest $\sigma$ indicated by blue triangles. As {\it Supersmoother} only returns values between 0 and 1 when it calculates the periodogram --- and thus the values are not normally distributed --- $\sigma$ as a statistic is meaningless by itself. We instead want to estimate the false-alarm probability (FAP) of each peak. Traditional estimates of significance (see \citealt{horne86}, for example) assume that the null hypothesis is pure white noise. Because the background noise is dependent on the period, we split the grid of periods into $N_{trial} = 100$ bins with 20 periods each. In each bin, we find the period $T$ associated with the largest value of $P_S$. We then calculate the number of simulated periodograms that have at least one point with power greater than $P_S(T)$ ($N_{DRW}(>P_S(T))$) within the period bin. The FAP is thus $N_{DRW}(>P_S(T))$ divided by the number of simulated DRW periodograms ($N_{DRW} = 96,000$) times $N_{trial}$,  which accounts for the fact that there are $N_{trial}\times N_{DRW}$ `chances' to randomly generate a peak with more power than the true peak (the look-elsewhere effect). 

\subsection{Distinguishing Periodicity from Systematics}

The above process results in a number of periods that correspond to local minima in FAP vs. $T$, shown as blue triangles in the right side of Figure \ref{fig:periodogram}. Between the sampling of the lightcurve and the algorithm used to generate the DRW lightcurves, it is possible that some of these detections are only arising due to artificial suppression of the DRW noise. To determine this, we use the same algorithm to simulate white noise lightcurves ($\tau\rightarrow 0$, with $\sigma$ and $\langle g\rangle$ drawn from the DRW samples in Figure \ref{fig:drw_samples}), and calculate the average ($P_{WN}$) and standard deviation ($P_{\sigma,WN}$) of the periodograms. $P_{WN}\pm P_{\sigma,WN}$ is shown in purple on the left panel of Figure \ref{fig:periodogram}. It is clear that $P_{WN}$ and $P_{\sigma,WN}$ are roughly constant over the range of tested periods, and thus that none of the detected periodicities arise due to suppression of the DRW noise.

\begin{figure*}[ht!]
\plotone{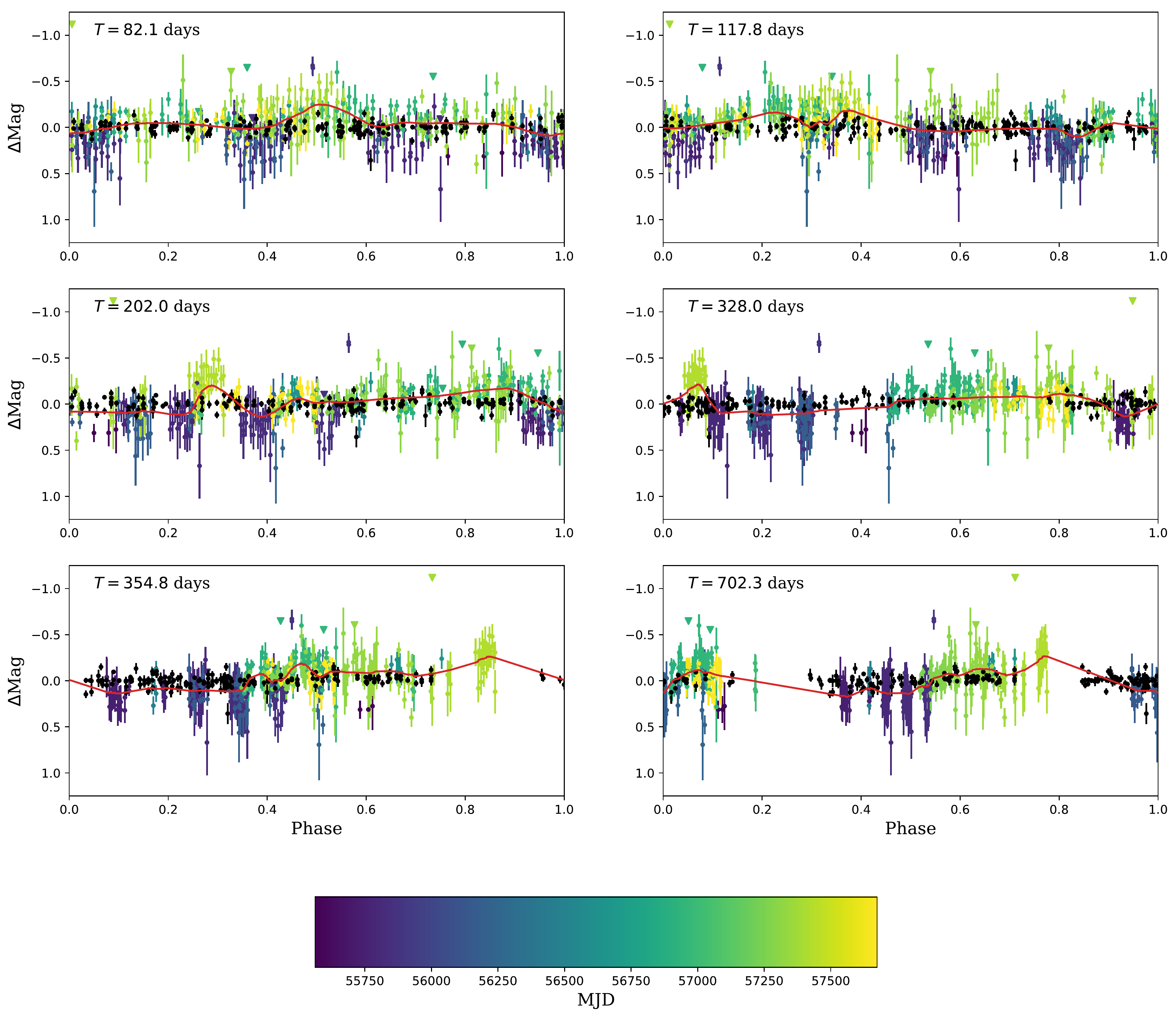}
\caption{Phase-folded, mean-subtracted data and best-fit {\it Supersmoother} model in red, for each of the seven periods detected with {\tt celerite}, as well as the $r$-band data in black. The $g$-band data are colored by the MJD of the observation. This coloring shows that multiple cycles of the period are observed, and the cycles are largely consistent. Upper limits for any non-detections are labeled with downward pointing triangles.\label{fig:period_model}}
\end{figure*}

It is also possible that the period detected at $T = 354.8$ days is due to approximately yearly systematic variations in observing conditions --- e.g., airmass, observability, weather, etc. --- at Palomar Observatory, and that the period at $T = 708.5 \approx 2\times354.8$ is an alias of the same effects. This appears to be reflected in Figure \ref{fig:period_model}, where the phase-sampling of both the $g$ and $r$ band data is nearly identical at these periods. Because J0045+41 is nearly at the detection limit of PTF, it is certainly possible that those systematics can masquerade as real effects; our discussion of these results comes with the major caveat that the $\sim$yearly periodicity may not be real. However, even discounting the $354.8$-day period, there is a secondary peak at $328$ days that is unlikely to be a result of these yearly systematics. 

\begin{figure}[ht!]
\plotone{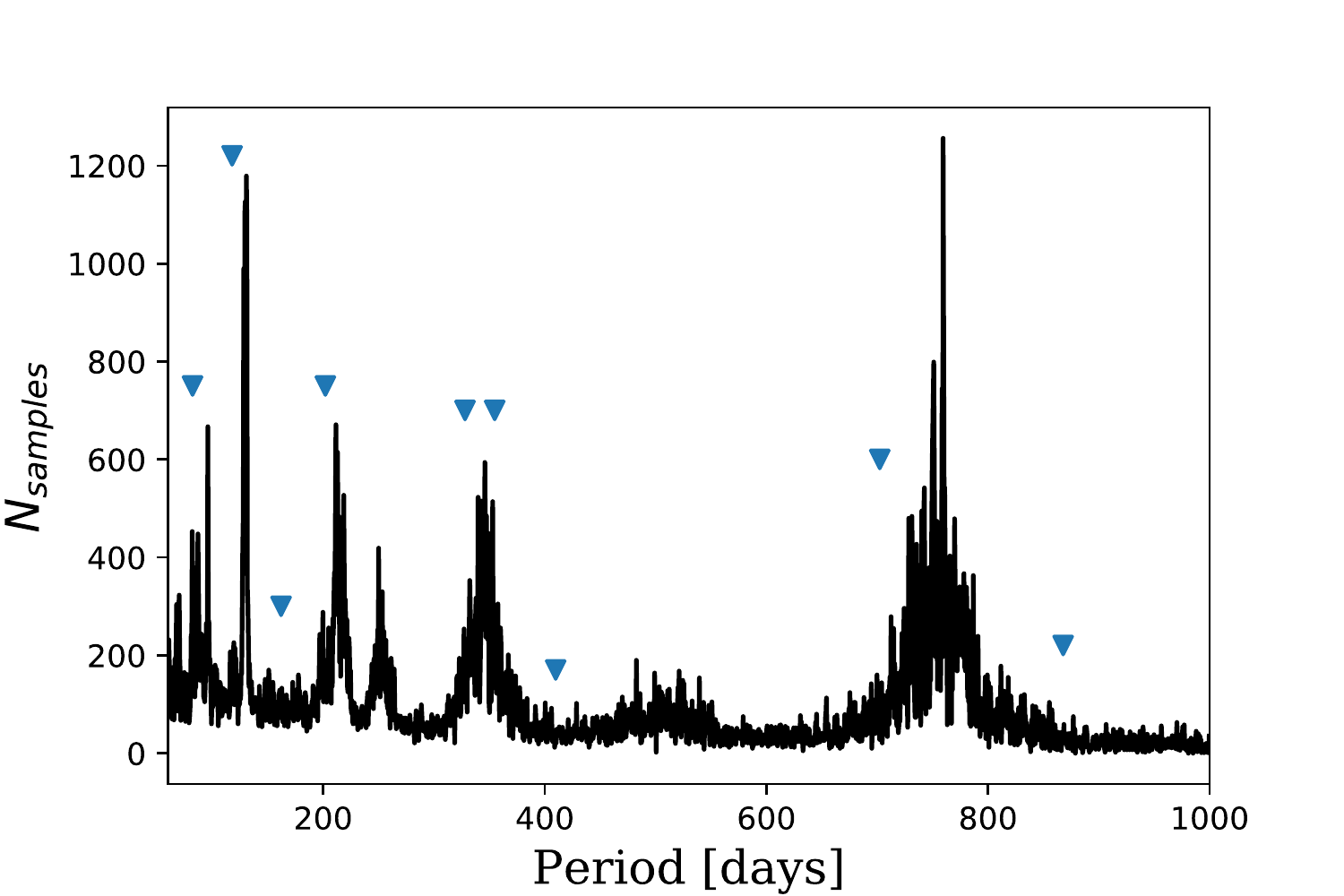}
\caption{Posterior distribution of periods, sampled by {\tt emcee}. Blue triangles indicate the ten peaks identified in $\sigma(T)$.}\label{fig:periodic_samples}
\end{figure}

Finally, if these periods are real, they should be detectable by other means. We add a sinusoidal mean model to our implementation of the DRW kernel within {\tt celerite}, and simultaneously sample the posterior distribution of the model parameters --- mean, amplitude, period, and phase --- and the DRW parameters as described above using {\tt emcee}, using double the number of walkers, and restricting the period of the sinusoid to lie between 60 and 1000 days. As discussed above, a sinusoidal model is not necessarily an accurate one; however, the periods revealed by this analysis should be similar to the periods found above. A histogram of the posterior distribution of the period is shown in Figure \ref{fig:periodic_samples}, with the periods with local minima in FAP indicated by blue triangles. It is clear that at least some of the peaks found --- namely at $T = 82.1,117.8,202.0,328.0,354.8,\:{\rm and},708.3$ days --- are retrieved. The phase-folded, mean-subtracted data and the best-fit {\it Supersmoother} model at the six periods detected with {\tt celerite}, along with the phase-folded $r$-band are shown in Figure \ref{fig:period_model}. Table \ref{tab:supersmoother_results} contains the period $T$, the value of $P_S(T)$, the bounds of the period bin containing $T$ ($T_{min}$ and $T_{max}$), the estimated FAP, and whether a strong peak in the {\tt celerite} posterior appears at a similar period.

\begin{deluxetable*}{lcccc}
\tabletypesize{\scriptsize}
\tablecaption{Results from \S\ref{sec:period}. $T$ is the period, $P_S(T)$ is as described in the text, $T_{min}$ and $T_{max}$ are the bounds of the period bin in which the FAP is calculated. The last column shows whether the period is detected using a DRW + sinusoidal mean model in {\tt celerite}.}\label{tab:supersmoother_results}
\tablehead{\colhead{$T$} & \colhead{$P_S(T)$}  & \colhead{($T_{min},T_{max})$} & \colhead{FAP} & \colhead{Detected with} \\
\colhead{days} & \colhead{} & \colhead{days}  & \colhead{} & \colhead{{\tt celerite}?}
}
\startdata
82.10 & 0.120592  & (78.809,79.280) & $6.98469 \times 10^{-3}$ & Yes \\ 
117.84 & 0.139525  & (116.428,116.898) & $7.14281 \times 10^{-3}$ & Yes \\ 
162.04 & 0.148967  & (154.047,154.517) & $7.78917 \times 10^{-3}$ & No \\ 
202.01 & 0.212229  & (201.071,201.541) & $4.72885 \times 10^{-3}$ & Yes \\ 
328.03 & 0.233829  & (323.332,323.802) & $3.30188 \times 10^{-3}$ & Yes \\ 
354.84 & 0.270498  & (351.546,352.016) & $1.01854 \times 10^{-3}$ & Yes \\ 
409.85 & 0.248934  & (407.974,408.444) & $4.27479 \times 10^{-3}$ & No \\ 
702.34 & 0.281859  & (699.520,699.990) & $4.59042 \times 10^{-3}$ & Yes \\ 
867.86 & 0.300183  & (859.400,859.870) & $5.84198 \times 10^{-3}$ & No \\ 
\enddata
\end{deluxetable*}

\begin{figure}[ht!]
\plotone{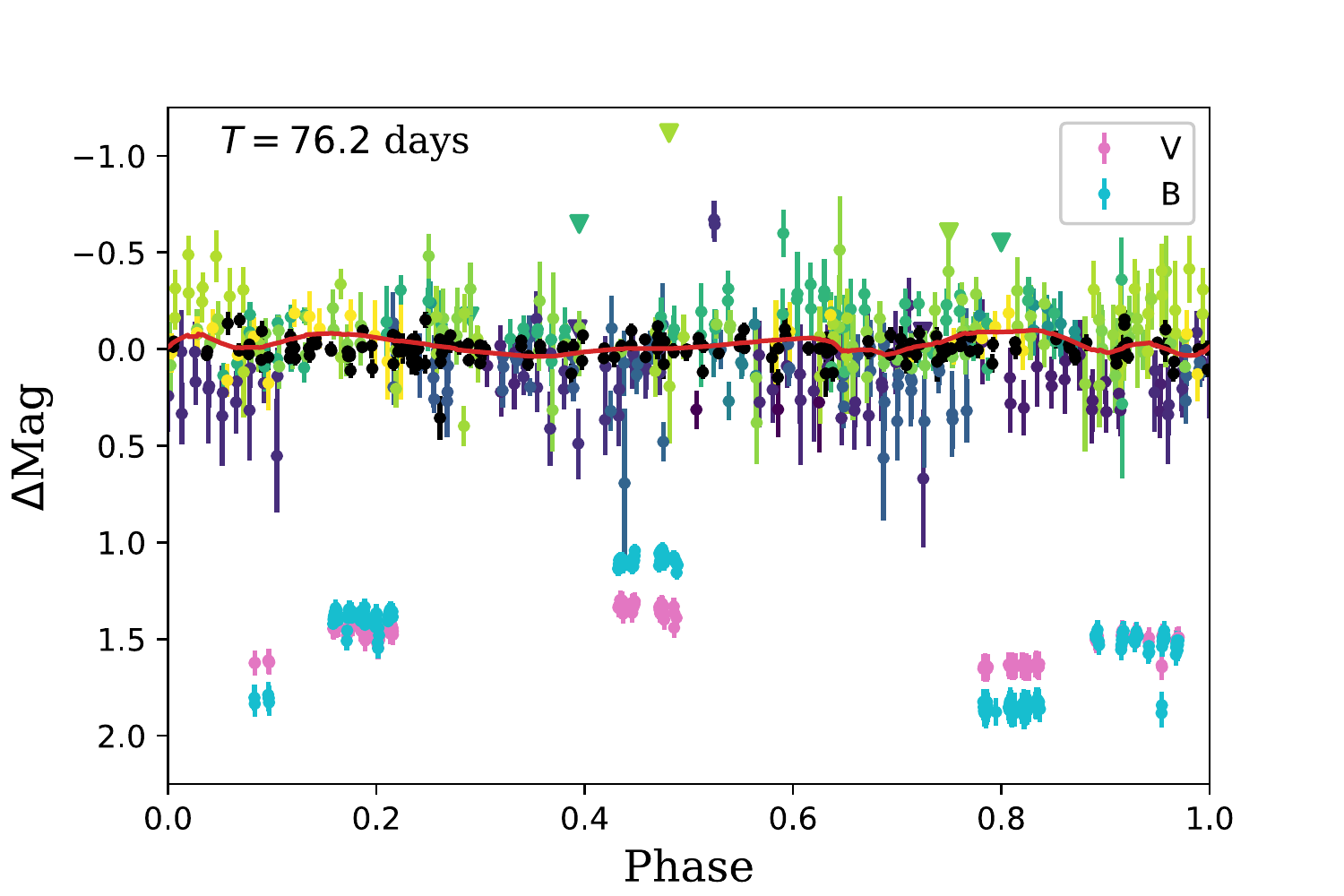}
\caption{Similar to Figure \ref{fig:period_model}, phase-folded on the period detected by \citet{vilardell06}. The data from \citet{vilardell06} are shown in blue and pink, and offset by a constant for clarity.}\label{fig:vil_period_model}
\end{figure}

The period of $\sim82.1$ days (FAP$\sim0.007$) is similar to \citet{vilardell06} who find a period of $\sim76$ days. We plot the PTF data, the historical data from \citet{vilardell06} (offset by a constant for clarity), and the best-fit {\it Supersmoother} model folded on the period found by \citet{vilardell06} in Figure \ref{fig:vil_period_model}. None of the structure in the \citet{vilardell06} data is seen in the PTF data or the {\it Supersmoother} fit; however, with so few observations, it is possible that the true period detected by \citet{vilardell06} is closer to that detected in the PTF data. Unfortunately, the historical data are only available phase-folded, and we are unable to include them in our analysis of other periods. 

\section{Discussion and Conclusion} \label{sec:discuss}

One possible interpretation of a periodic signal in an AGN is that it is due to the orbital motions of a SMBHB, formed through a major galaxy merger. Though small, the number of $z < 1$ candidate SMBHBs discovered is consistent with this model \citep{volonteri09}. The detected periodicities of J0045+41 are thus quite interesting. Most intriguingly, the $\sim82.1$ day period is almost exactly in a 1:4 ratio with the $\sim328$-day period. It is possible that either of these peaks is an alias of the other, as the observed periodogram is the convolution of the true periodogram with the Fourier transform of the sampling function \citep{roberts87,charisi15}. However, multiple periodicities beyond the orbital period are predicted to occur in SMBHBs at similar period ratios as a result of interactions in the circumbinary disk \citep{macfadyen08,shi12,farris14}. 

In particular, \citet{macfadyen08} found that the periodogram of the accretion rate in their simulation displayed significant peaks at frequencies approximately generated by the formula $\omega = \frac{2}{9}K\Omega_{bin}$, where $\Omega_{bin}$ is the binary orbital angular frequency, and $K = 1,2,6,7,8,9,10$. We search for the orbital period, $T_{bin}$, that generates a set of periods closest to the first five observed periods (discounting the 354.8 and 708.5 day periods). We find that $T_{bin} = 169.29$ ($\Omega_{bin} = 3.7\times10^{-2}$ day$^{-1}$) creates periods that match quite well with the two shortest periods, though it underpredicts the 202 day period by $\sim75$ days, and overpredicts the 328 day period by $\sim50$ days. Finally, \citet{farris14} find that, for varying mass ratios and simulation setups, periodic variations in the accretion rate onto one or both black holes can arise at frequencies with the same 1:4 correspondence as the 82 and 328 day periods. These occur at $1/4\Omega_{bin}$ and $\Omega_{bin}$. This points to the 82.1 day period being the orbital period of the binary. \citet{farris14} also find frequencies arising at $\Omega_{bin}$ and $2\Omega_{bin}$. Interestingly, we do detect a period with FAP$\sim0.008$ at $162\approx2\times82\approx\frac{1}{2}\times328$ days. While we do not detect a strong peak in the {\tt celerite} posterior around this period, this hints that the orbital period may also be 162 or 328 days.

If we assume that any of these three periods is the orbital period of a SMBHB in a circular Keplerian orbit, and that the virial mass derived in \S\ref{sec:spec} is the total mass of the two black holes $M_{tot}$, then the semimajor axis of the orbit ranges from 216 to 544 AU (or 0.3 to 1 microarcseconds at the angular diameter distance of J0045+41, which is unresolvable using current radio interferometric arrays). Such a separation would be well within the regime where loss due to gravitational radiation is significant. We can approximate the time for two circularly orbiting black holes to inspiral due to gravitational radiation using equations (5.9) and (5.10) from \citet{peters64}:
\begin{equation}
\begin{array}{ccc}
t_{GW} & = & \frac{5}{256}\frac{c^5}{G^3}\frac{R^4}{(M_1 + M_2)(M_1M_2)}  \\
& = & \frac{5}{256}\frac{c^5}{G^3}\frac{R^4}{M_{tot}^3}\frac{(1+q)^2}{q}
\end{array}
\end{equation}
where $R$ is the semimajor axis of the orbit, $M_1,\:M_2$ are the masses of the individual black holes and $q \equiv M_2/M_1$. $t_{GW}$ ranges between $\sim350$ yr (for the shortest period, with $q = 1$) to 360 kyr (for the longest period, with $q = 0.01$). 

The gravitational waves produced by SMBHBs are expected to be detectable at the nHz frequencies probed by pulsar timing arrays (PTAs, \citealt{foster90}). The amplitude of the dimensionless gravitational strain ($h_0$) of a SMBHB with mass ratio $q$ at redshift $z$, assuming a circular orbit with period $T$ can be expressed as
\begin{equation}
h_0 = \frac{4G}{c^2}\frac{qM_{tot}}{(1+q)^2D_L(z)}\bigg(\frac{2\pi GM_{tot}}{c^3 T}\bigg)^{2/3}
\end{equation}
where $D_L(z)$ is the luminosity distance \citep{thorne87}. The expected strain of a SMBHB with the derived mass and orbital period of J0045+41 would range from $\sim10^{-16}$ (for the shortest detected period, with $q = 1$) to $\sim10^{-18}$ (for the longest period, with $q = 0.01$). These results, in addition to the expected orbital velocity of the secondary black hole (see below) are summarized in Table \ref{tab:orbit_results}. While the latter strain would be orders of magnitude below the stochastic background of gravitational radiation from all SMBHBs at that period ($h \approx 10^{15}$ at $T = 1$ yr, \citealt{shannon13}), the background falls off at higher frequencies as fewer sources are expected to be inspiraling at shorter and shorter periods, and the signal from a $\sim80$ day SMBHB would be detectable above the background \citep{moore15}. Indeed, the signal would be just shy of the anticipated sensitivity --- $\sim6\times10^{-16}$ \citep{lazio13} --- of the Square Kilometer Array (SKA, \citealt{dewdney09}). While this is an exciting finding, it is important to note that there are a number of other possible interpretations of a periodic signal, e.g: a long-lived or periodically-generated hot spot in the accretion disk, geodetic precession, and self-warping of the disk (see \citealt{bon17} for a concise review). 

\begin{deluxetable*}{lcccc}
\tabletypesize{\scriptsize}
\tablecaption{Orbital and gravitational properties of proposed orbital periods}\label{tab:orbit_results}
\tablehead{\colhead{$T$} & \colhead{$R$/$\theta$} & \colhead{$v_{orb}$} & \colhead{$t_{GW}$} & \colhead{$h_0$} \\
\colhead{days} & \colhead{AU/$\mu$arcsec} & \colhead{$10^3$ km s$^{-1}$}  & \colhead{yr} & \colhead{}\\
\colhead {}    & \colhead{}               & \colhead{($q = 1/0.01$)} & \colhead{($q = 1/0.01$)} & \colhead{($q = 1/0.01$)}
}
\startdata
82.10 & 216.02/0.30 & 14.312/28.341 & $3.522 \times 10^{2}$/$8.982 \times 10^{3}$ & $9.252 \times 10^{-17}$/$3.628 \times 10^{-18}$  \\ 
162.04 & 339.90/0.47 & 11.410/22.594 & $2.159 \times 10^{3}$/$5.505 \times 10^{4}$ & $5.880 \times 10^{-17}$/$2.306 \times 10^{-18}$  \\ 
328.03 & 543.93/0.75 & 9.020/17.860 & $1.416 \times 10^{4}$/$3.610 \times 10^{5}$ & $3.674 \times 10^{-17}$/$1.441 \times 10^{-18}$  \\ 
\enddata
\end{deluxetable*}

Even if it is not a SMBHB, J0045+41 is an interesting object. For one, it appears to be probing a relatively extinction-free region of the ISM in M31. The detection of the \ion{Na}{1} D doublet is promising, and follow-up optical and infrared observations at higher spectral resolution may disentangle absorption from M31 and from the Milky Way, and reveal more about the dynamics of the ISM along the line of sight towards J0045+41. The spectrum is well fit by a mixture of the galaxy and quasar eigenspectra from \citet{yip04,yip04b} redshifted to $z = 0.215$ and reddened by an $A_V = 1.0\pm1.0$ mag \citet{cardelli89} extinction law. However, H$\alpha$ and H$\beta$ both have a blueshifted broad component. Indeed, the residuals to the fit shown in Figure \ref{fig:fit} appear to be Gaussian. Fitting these residuals with a Gaussian profile shows that this component is at $z = 0.196$, a $\sim4800$ km s$^{-1}$ difference from the host redshift. This shift may be due to an outflow from the central engine, a hot spot in the blueshifted side of the accretion disk, or the blending of the broad lines of each SMBH component; as the less massive SMBH moves towards us, we would see its broad lines blueshifted, which would explain the excess of blue flux in the broad lines \citep{shen10}. Indeed, a similar binary model has been used to explain SDSS J092712.65+294344.0, which also appears to have blueshifted broad lines relative to the narrow lines in the spectrum \citep{dotti09,bogdanovic08}. At the short periods found in \S\ref{sec:period}, orbital velocities are expected to be $\sim10^4$ km s$^{-1}$ (depending on the assumed mass ratio), so this blueshift would be consistent with the orbital velocities for all of the periods in Table \ref{tab:orbit_results}, for any value of the mass ratio. Follow-up spectroscopy on a cadence of a few months would be able to search for or exclude periodic changes of the H$\alpha$ and H$\beta$ profiles relative to the narrow lines over time, which would help point to an explanation.

\begin{figure*}[ht!]
\plotone{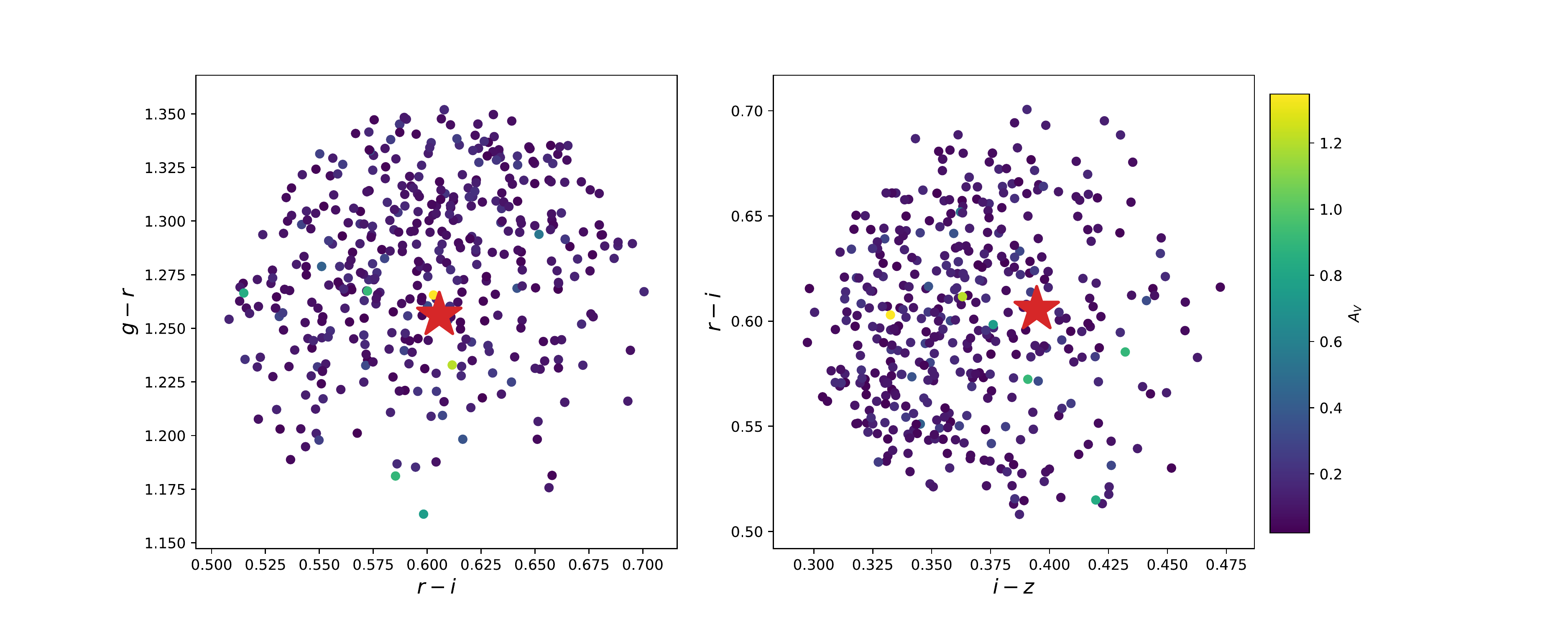}
\caption{$g-r$ vs. $r-i$ (left) and $r-i$ vs. $i-z$ (right) color-color plots of the SDSS $z<1$ quasar sample within 0.1 magnitudes of J0045+41, colored by assumed extinction in $g$. The red star is J0045+41.\label{fig:nearby}}
\end{figure*}

To search for any objects similar to J0045+41 in color space, we used PySynphot (a Python implementation of Synphot distributed by Space Telescope Science Institute, \citealt{lim15}) to generate synthetic photometry from our spectrum in $g$, $r$, $i$, and $z$ --- there was not enough signal in $u$ to synthesize a magnitude. We then downloaded photometry of all low-redshift ($z<1$) SDSS quasars from Data Release 13 \citep{sdss16} within 0.1 magnitudes of J0045+41 in $g-r$ vs. $r-i$ vs. $i-z$ color space. These quasars are shown in color space in Figure \ref{fig:nearby}. Each point is colored by the assumed value of the extinction in $g$. Of these 446 objects, only 197 of them have redshifts that are positive --- implying the remaining objects are not plausibly quasars. Indeed, the spectra of many of the `quasars' in this sample are quite clearly cool stars. Some of these objects are simply misidentified; however, many are flagged with a {\tt Z\_WARNING: NOT\_QSO} by the SDSS pipeline. While this is helpful for reducing contamination of the quasar sample, it illustrates than many objects of interest fall through the cracks of classification algorithms (see \citealt{dornwallenstein17} for further discussion).

Of the true quasars in the sample, none are extincted by more than 1.5 magnitudes in $g$. It is likely that these quasars (and the AGN component of J0045+41) are {\it intrinsically} red as described by \citet{richards03}. These quasars may have been reddened by dust intrinsic to the host galaxy, or have excess red flux due to synchrotron emission with an optical turnover. Higher resolution spectroscopic follow-up would allow for more detailed fitting of J0045+41 to determine if a red quasar template yields a better fit.

The confusion of stars and quasars represents a unique problem for purely photometric surveys, such as the upcoming LSST project \citep{ivezic08}. Stars and higher redshift ($z>2.2$) quasars are well separated in color space. However, at lower redshifts, the two color loci appear closer and closer. The difference between the two populations is most apparent in $u$-band flux and $u-g$ color; indeed the $u$ filter was designed in part to leverage the difference between power-law spectra and spectra with strong Balmer decrements \citep{fukugita96,stoughton02,richards02}. Thus, in any single-visit catalog, the colors of the lower-redshift, low-luminosity, and intrinsically red AGN are the hardest population to distinguish from stars. LSST will visit most of its survey area $\sim50$-180 times in each filter over 10 years. \citet{peters15} demonstrated that it is possible to use variability in addition to colors to distinguish stars from AGN with high ($>90\%$) completeness. However, the accuracy of classifications in the lowest redshift bins studied drops to $\sim80\%$. While the number of quasars at low redshift is small, this highlights the importance of developing accurate classification algorithms for objects similar to J0045+41. Forthcoming work will focus on distinguishing between stars and quasars in the low-redshift, low-luminosity, red regime.

J0045+41 is an exciting and unique object. It represents an extreme end of color space in which photometric classification methods fail. Both the simple selection methods (described in \S\ref{sec:intro}) and more sophisticated machine learning algorithms are unable to correctly classify objects in this regime. Finding these intrinsically red AGN is important, as they are still poorly understood. The evidence of multiple periodic signals in the photometric lightcurve of J0045+41 is compelling, and warrants more dedicated spectroscopic observations at higher spectral resolution and deeper photometric observations sampled at a higher rate. Such observations would be crucial to confirm the presence of a SMBHB in J0045+41. They would also allow for the confirmation of the periods that we detected. The photometric data will soon be attainable in the form of the Zwicky Transient Facility (ZTF, \citealt{bellm14}), a next-generation transient survey that will see first light this year. 

\acknowledgments

The authors thank Jessica Werk, Julianne Dalcanton, Ben Williams, Jake VanderPlas and Scott Anderson for their valuable advice and feedback on this work. We wish to thank the anonymous referee for their extremely helpful comments. 
Based on observations (Program ID GN-2016A-FT-30) obtained at the Gemini Observatory (processed using the Gemini IRAF package), which is operated by the Association of Universities for Research in Astronomy, Inc., under a cooperative agreement with the NSF on behalf of the Gemini partnership: the National Science Foundation (United States), the National Research Council (Canada), CONICYT (Chile), Ministerio de Ciencia, Tecnolog\'{i}a e Innovaci\'{o}n Productiva (Argentina), and Minist\'{e}rio da Ci\^{e}ncia, Tecnologia e Inova\c{c}\~{a}o (Brazil). The authors thank the Gemini-North support staff. The scientific results reported in this article are based in part on data obtained from the Chandra Data Archive and the SDSS archive. Funding for the Sloan Digital Sky Survey IV has been provided by
the Alfred P. Sloan Foundation, the U.S. Department of Energy Office of
Science, and the Participating Institutions. SDSS-IV acknowledges
support and resources from the Center for High-Performance Computing at
the University of Utah. The SDSS web site is www.sdss.org.

SDSS-IV is managed by the Astrophysical Research Consortium for the 
Participating Institutions of the SDSS Collaboration including the 
Brazilian Participation Group, the Carnegie Institution for Science, 
Carnegie Mellon University, the Chilean Participation Group, the French Participation Group, Harvard-Smithsonian Center for Astrophysics, 
Instituto de Astrof\'isica de Canarias, The Johns Hopkins University, 
Kavli Institute for the Physics and Mathematics of the Universe (IPMU) / 
University of Tokyo, Lawrence Berkeley National Laboratory, 
Leibniz Institut f\"ur Astrophysik Potsdam (AIP),  
Max-Planck-Institut f\"ur Astronomie (MPIA Heidelberg), 
Max-Planck-Institut f\"ur Astrophysik (MPA Garching), 
Max-Planck-Institut f\"ur Extraterrestrische Physik (MPE), 
National Astronomical Observatories of China, New Mexico State University, 
New York University, University of Notre Dame, 
Observat\'ario Nacional / MCTI, The Ohio State University, 
Pennsylvania State University, Shanghai Astronomical Observatory, 
United Kingdom Participation Group,
Universidad Nacional Aut\'onoma de M\'exico, University of Arizona, 
University of Colorado Boulder, University of Oxford, University of Portsmouth, 
University of Utah, University of Virginia, University of Washington, University of Wisconsin, 
Vanderbilt University, and Yale University. This work was facilitated though the use of advanced computational, storage, and networking infrastructure provided by the Hyak supercomputer system at the University of Washington.

This work made use of the following facilities and software:

\vspace{5mm}
\facility{Gemini-North (GMOS)}

\software{aplpy v1.1.1 \citep{robitaille12}, Astropy v1.2.1 \citep{astropy13}, Ciao v4.7 \citep{fruscione06}, celerite v0.3.0 \citep{foremanmackey17}, corner v2.0.1 \citep{foremanmackey16}, emcee v2.2.1 \citep{foremanmackey13}, gatspy v0.3 \citep{vanderplas15,vanderplascode15}, IRAF v2.16 \citep{tody86}, Matplotlib v2.0.0 \citep{hunter07}, NumPy v1.11.3 \citep{vanderwalt11}, PySynphot v0.9.8.4 \citep{lim15}, and Sherpa v4.7 \citep{freeman01}  
          }

\end{document}